**Title** Response Surface Methodology coupled with desirability functions for multi-objective optimization: minimizing indoor overheating hours and maximizing useful daylight illuminance

**Authors:** * Gamero-Salinas, Juan [a] [b] (jgamero@unav.es); López-Fidalgo, Jesús [a] [b] (fidalgo@unav.es)

**Affiliation:**
[a] University of Navarra. Institute of Data Science and Artificial Intelligence (DATAI), University Campus, 31008 Pamplona, Spain
[b] University of Navarra. TECNUN School of Engineering, Manuel Lardizábal 13, 20018, San Sebastián, Spain

**Abstract**

Response Surface Methodology (RSM) and desirability functions were employed in a case study to optimize the thermal and daylight performance of a computational model of a tropical housing typology. Specifically, this approach simultaneously optimized Indoor Overheating Hours (IOH) and Useful Daylight Illuminance (UDI) metrics through an Overall Desirability (D). The lack of significant association between IOH and other annual daylight metrics enabled a focused optimization of IOH and UDI. Each response required only 138 simulation runs (~30 hours for 276 runs) to determine the optimal values for passive strategies: window-to-wall ratio (WWR) and roof overhang depth across four orientations, totalling eight factors. First, initial screening based on $2_V^{8-2}$ fractional factorial design, identified four key factors using stepwise and Lasso regression, narrowed down to three: roof overhang depth on the south and west, WWR on the west, and WWR on the south. Then, RSM optimization yielded an optimal solution (roof overhang: 3.78 meters, west WWR: 3.76%, south WWR: 29.3%) with a D of 0.625 (IOH: 8.33%, UDI: 79.67%). Finally, robustness analysis with 1,000 bootstrap replications provided 95% confidence intervals for the optimal values. This study optimally balances thermal comfort and daylight with few experiments using a computationally-efficient multi-objective approach.

# 1. Introduction

This section reviews how the mitigation of indoor overheating and the improvement of daylight performance are being studied in tropical dwellings, with a focus on the use of Response Surface Methodology (RSM) for multi-objective optimization of these responses, concluding with the identification of a research gap.

## 1.1. Background

The tropical region is projected to account for over half by 2050 [1–3]. Over the last century, the Tropics have warmed by 0.7–0.8°C, with models predicting an additional 1–2°C rise by 2050 [4]. In Latin America and the Caribbean (LAC), increasing temperatures and humidity have driven higher demand for energy services, particularly for cooling. For example, air conditioning usage in Honduras has surged in its largest cities; in Tegucigalpa, it rose from 3.9% in 2001 to 9.3% in 2013, while in San Pedro Sula, it jumped from 11.6% to 26.9% [5]. However, progress on building codes for thermal comfort and energy efficiency has been slow. By 2018, only six LAC countries had adopted mandatory or voluntary energy codes, with none in Central America [6–8]. Energy efficiency efforts in Honduras, like the 2024 Law for Rational and Efficient Energy Use and the 2019 Sustainable Construction Guide for Tegucigalpa, have made progress but lack specific targets for indoor thermal comfort and daylight [9,10]. Urban areas in tropical regions face intense heat stress, leading to a high risk of building overheating [11]. However, passive strategies—which by definition do not require the aid of mechanical means—can reduce overheating and limit reliance on air conditioning. Yet, improving thermal comfort often impacts daylight levels, creating a trade-off between thermal and daylight performance that remains underexplored, particularly in warm-humid tropical dwellings. This will be addressed in subsections 1.2 and 1.3.

## 1.2. Overheating mitigation and daylight performance

Typically, multi-objective optimization for thermal comfort and daylight is focused in office, commercial, and educational settings, with limited research in warm-humid tropical residential contexts [12–18]. To the best of the authors' knowledge, only two studies have simultaneously assessed the impact of passive measures on thermal comfort and daylight performance in tropical dwellings, and neither addresses thermal comfort specifically in terms of 'overheating' [19,20]. The existing literature on tropical dwellings (using both air conditioning and/or natural ventilation) treats overheating (in terms of indoor overheating hours or hours of exceedance) as a separate issue from daylight performance, where emphasis is given to passive cooling measures such as solar shading/protection (e.g. brise-soleil, balconies, roof overhangs) on windows, window length and height or window-to-wall ratio (WWR), orientation, natural ventilation, roof and wall properties (e.g. thermal absorptance, insulation), and glass solar heat gain coefficient [21–27]. In contrast, studies on daylight performance in tropical contexts do not address indoor overheating or thermal comfort [28–33]. These studies typically use annual daylight metrics such as Useful Daylight Illuminance (UDI), Annual Sunlight Exposure (ASE), Daylight Autonomy (DA), or Spatial Daylight Autonomy (sDA), or image-based metrics like Daylight Glare Probability (DGP), or point-in-time metrics like Daylight Factor (DF), which is calculated under an unobstructed overcast sky. In these studies, emphasis is given to passive daylight measures such as WWR, orientation, indoor wall reflectance or orientation of interior partitions.

## 1.3. Response Surface Methodology

Design of Experiments (DoE) is especially important in practice when an experiment is costly, time-consuming, or ethically challenging [34,35]. The RSM, a specific technique within the broader field of DoE, is commonly used in engineering, chemistry, and other applied sciences, and is a collection of statistical and mathematical techniques useful for developing, improving, and optimizing processes [35–37]. *However, its use in the building performance industry is limited.* Table 1 summarizes the studies found in the literature where RSM has been applied in this field. Throughout the text, the term 'design' specifically refers to the context of DoE not to passive cooling/daylight design or architectural/building design.

| Ref. | Experiment setup | Inputs | Responses | Design of Experiments |
|---|---|---|---|---|
| [38] | EnergyPlus & DAYSIM | Window geometry | - Energy consumption and daylight (i.e. DF and DA) | RSM is used, but design is not specified |
| [39] | Survey + monitoring | Indoor environmental quality factors (i.e. air temperature, relative humidity, noise, $CO_2$, VOCs) | - Occupant productivity | RSM is used, but design is not specified |
| [40] | EnergyPlus | HVAC, envelope, internal loads | - Thermal comfort (i.e. time outside thermal comfort zone) | RSM is used, but design is not specified |
| [41] | EnergyPlus | Double-skin façade factors (e.g. slat angle, cavity depth) | - Thermal performance (i.e. total heat gain density) | - Definitive Screening Designs<br>- Plackett-Burman Design<br>- Taguchi 2-Level Designs<br>- Central Composite Design<br>- Box-Behnken Design |
| [42] | Lab-type experiment (i.e. climate chamber) | Double-skin façade factors (e.g. venetian blind angle, temperature difference, opening size) | - Thermal performance (i.e. net heat flux density, air temperature, average surface temperature of the cavity) | - Definitive Screening Designs<br>- Taguchi 2-Level Designs<br>- Central Composite Design |
| [43] | Computational Fluid Dynamics (CFD) simulation (ANSYS Fluent) | Wind speed and direction | - Ventilation rate in $m^3/s$ | - Central Composite Design<br>- Full Factorial Design<br>- D-optimal designs |
| [44] | Computational Fluid Dynamics (CFD) simulation (ANSYS Fluent) | Window types, orientation, ventilation modes, wind speed and direction | - Ventilation rate (i.e. air changes per hour) | - Central Composite Design |
| [45] | Lab-type experiment (i.e. two-test rooms) | Phase-change material and ceiling fan ventilation factors (e.g. inlet air temperature, fan height, material thickness) | - Thermal comfort (i.e. percentage of people dissatisfied) | - Central Composite Design |
| [46] | EnergyPlus | Insulation thickness, type, location | - Total heating and cooling loads and total costs | - Box-Behnken Design |
| [47] | DesignBuilder/EnergyPlus | Heating and cooling setpoints, U-value | - Total energy demand in kWh | - Central Composite Design |

Table 1. Literature review using the RSM in the context of building performance.

Some studies have utilized RSM without conducting a prior DoE, meaning that the data used to characterize the response with surface models were not derived from a formal experimental design [38–40]. In contrast, most studies utilize RSM based on a DoE (e.g. Placket-Burman, Central Composite Design), but primarily for screening or characterization, rather than optimization [41–47]. To the author's knowledge, no study in the context of building simulation or lab-type experiments has focused on optimizing experimental factors using RSM alone; those that pursue optimization typically rely on evolutionary/genetic algorithms (e.g. NSGA-II, NSGA-III) [48–54].

### 1.4. Research gap and aim

Subsections 1.2 and 1.3 present a gap where the simultaneous optimization of overheating and daylight remains unexplored. Given this gap and the opportunity that RSM provides on optimizing responses with few experiments—thanks to its DoE foundation—and considering the computational demands of thermal and daylight simulations, *this research aims to use an RSM approach to identify optimal values for passive measures related to thermal and daylight comfort using as case study a tropical dwelling typology.* The goal is to minimize indoor overheating and maximize daylight performance while addressing the conflict between these responses: increasing fenestration and reducing shading can enhance natural light but may also result in increased thermal discomfort and glare.

## 2. Methodology

This section describes the methodology followed for the multi-objective optimization of thermal comfort and daylight is illustrated in Figure 1.

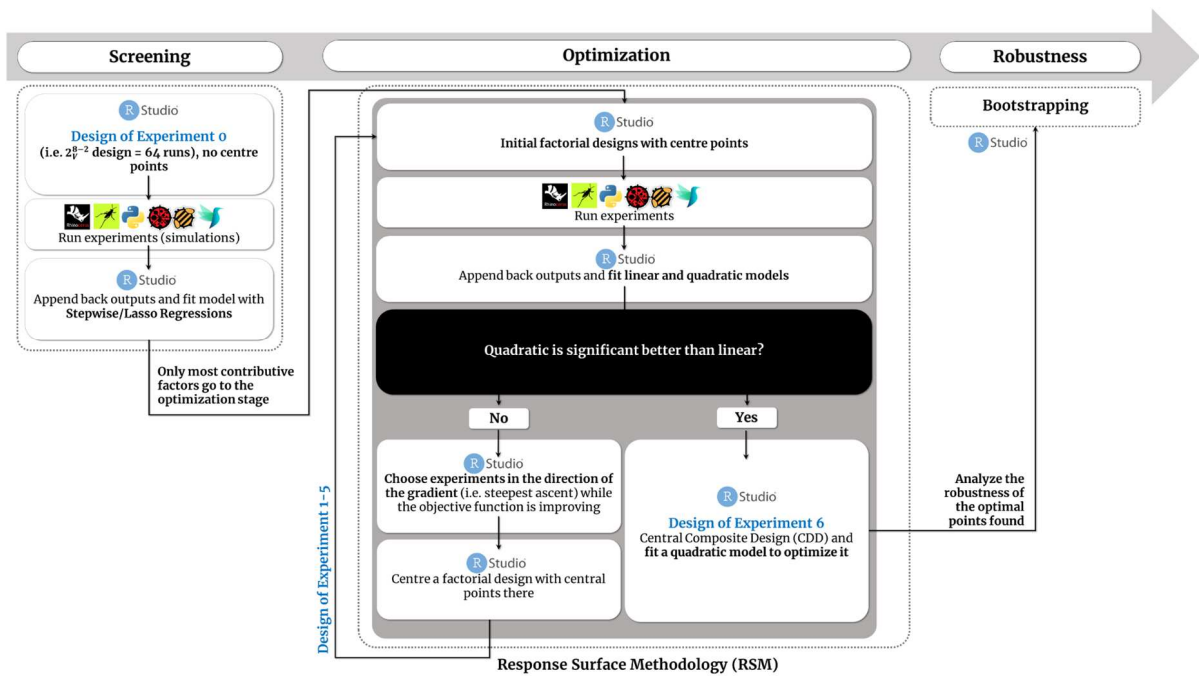

Figure 1. Flowchart of the methodology employed in this research study

### 2.1. Thermal comfort and daylight modelling

Dynamic simulations were carried out using the *Ladybug* v. 1.8.0 and *Honeybee* v. 1.0.0 tools [55] in *Grasshopper* within *Rhino* v.7, with thermal comfort simulations utilizing the *OpenStudio* v. 3.7.0 engine [56] and daylight simulations using the *Radiance* v. 1.0 engine [57]. The *Colibri* v. 2.0 component [58] was used to iterate the simulations parametrically through all the possible combinations given by both the *fractional factorial design* and the *response surface designs* explained in 2.3 and 2.4 subsections, respectively. Colibri compiles into a csv file both input and output values per iteration. Subsequently, upon completion output simulation values were appended back RStudio v. 2024.04.2+Build764 [59] for the screening and optimization stages, respectively.

*2.1.1. Building model*

San Pedro Sula (15.45º, -87.22º, 31m above sea) is a Honduran city classified as *Af* in the Koppen-Geiger climate classification and 0A in the ASHRAE 169-2020 climate classification, with an annual average dry bulb temperature of 27.2ºC [60]. As shown in Figure 2, San Pedro Sula experiences two major meteorological seasons: rainy season from March to October (hottest temperatures, highest humidity, strongest winds, and most solar radiation), and dry season from November to February (coolest temperatures, lowest humidity, lightest winds, and least solar radiation). This city has a tropical lowland climate, similar to Miami, Dhaka, Dar-es-Salaam, Mombasa or Havana [61]. The available weather file from Ladybug Tools (i.e. *HND_La.Mesa-San.Pedro.Sula.787080_SWERA.epw*) was used for running the thermal comfort and daylight simulations.

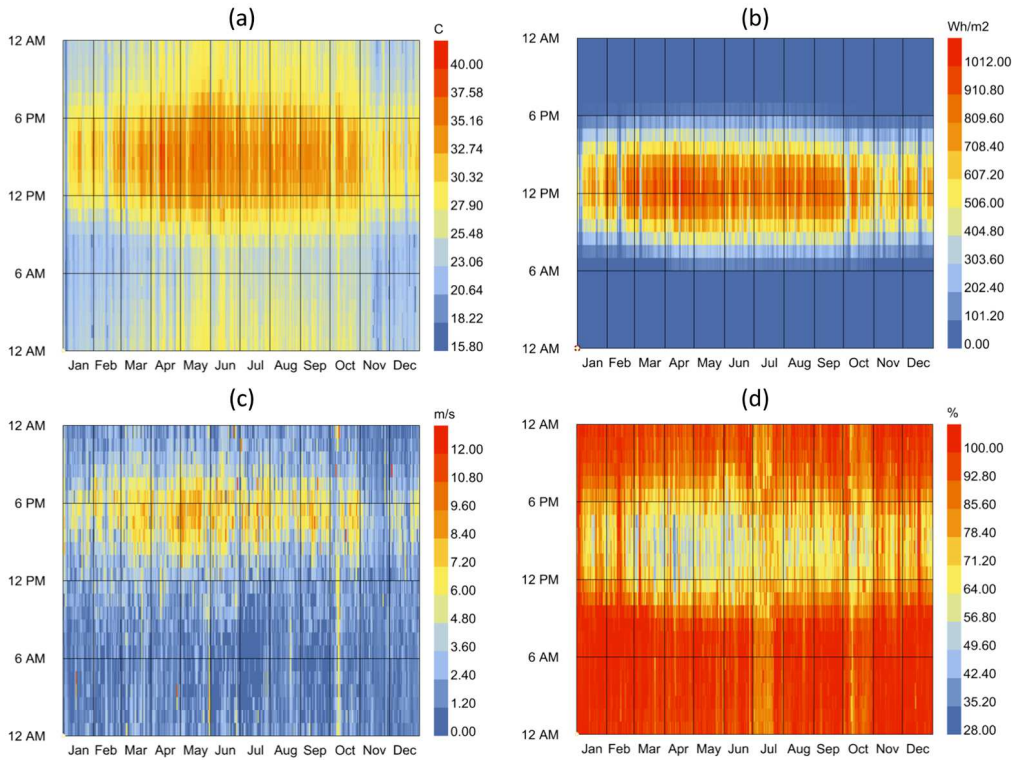

*Figure 2. Key outdoor weather parameters for San Pedro Sula derived from its Ladybug .epw file: (a) dry bulb temperature, (b) global horizontal radiation, (c) wind speed, and (d) relative humidity.*

In Honduras, approximately 85% of the currently built houses follow a single-family housing typology based on national historical data [62]. Although multi-family housing typology building have been increasing in the last years, the single-family housing typology remains the most representative one and it's still the primary approach employed by the different governments, and national and international NGOs to satisfy the Honduran housing deficit that as of 2022 goes up to almost 1.5 million housing units, including those affected by Eta and Iota Hurricanes of 2020 [63]. In this study a typical social housing typology was modelled, in a simplified way, using Rhino and Grasshopper as a 60 square-meter house, with its main facades facing north and south assuming this orientation as a best-case scenario for avoiding high indoor overheating and having maximum good illuminance levels. In this tropical context, the south facade receives direct solar radiation for the majority of the year, while the north receives it for 3 months (e.g., in San Pedro Sula approximately from May 3 – August 11). As shown in Figure 3. five thermal zones were modelled: open plan living-dining room (20.86m$^2$), the kitchen room (8.54m$^2$), bedrooms (12.45m$^2$), and WC (4.5m$^2$).

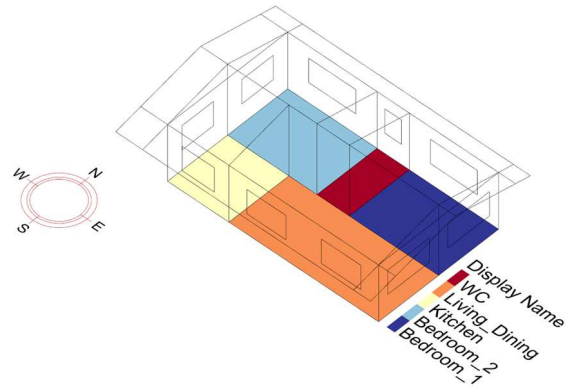

*Figure 3. Rhino/Honeybee visualization of the housing typology, at simulation run 77.*

The digital model of the dwelling typology was modelled using the values indicated in Table 2, which remained constant across all simulation runs. Additional model details are available in the Honeybee model file, provided in the Supplementary Materials section.

| Building construction factor | Description |
| --- | --- |
| Natural ventilation type | Wind-driven cross ventilation |
| Natural ventilation schedule | All rooms: Off (8-13h) & On (0-7h, 14-23h) |
| Operable window area | 50% |
| Window discharge coefficient | 0.45 |
| Glass thickness | 3mm |
| Window height | 1.2 meters |
| Window sill height | 1.0 meters |
| Glass solar radiation and visible light transmittance | 0.85 (for clear uncoated glass) |
| Glass solar radiation and visible light reflectance | 0.075 (for clear uncoated glass) |
| Exterior wall & roof solar absorptance | 0.30 (light colour) |
| Exterior wall & roof U-value | 2 W/m2K |
| Exterior wall thermal mass value | 130 kJ/m2K |
| Roof thermal mass value | 10 kJ/m2K |
| Air infiltration | 0.0006 m3/s per m2 façade (at a typical building pressure of ~4 Pa) |
| Lighting schedule | Bedrooms & WC: Off (0-5h, 8-21h) & On (6-7h, 22-23h) |
| | Living, Kitchen & Dining: Off (0-15h, 22-23h) & On (16-21h) |
| Lighting power density | Bedrooms & WC: 45 W/m2 |
| | Living, Kitchen & Dining: 60 W/m2 |
| Lighting load radiant fraction | Bedrooms & WC: 0.3 |
| | Living, Kitchen & Dining: 0.3 |
| Occupation schedule | Bedrooms & WC: Off (8-21h) & On (0-7h, 22-23h) |
| | Living, Kitchen & Dining: Off (0-13h, 22-23h) & On (18-21h, 14-17h 50% occupancy) |

*Table 2. Housing typology model fixed characteristics for energy and daylight simulations.*

The model assumed *wind-driven cross ventilation*, considering monitoring and simulation studies that highlight its effectiveness in reducing overheating risk [21,22]. This approach assumes pressure differences across windows on opposite sides of the dwelling, as is the case in the study scenario. The operable window area was set at 50%, reflecting the typical scenario for sliding windows. A discharge coefficient of 0.45 was applied, representing the typical value for unobstructed windows with insect screens, commonly found in Honduran social housing projects. Natural ventilation is shut off when the temperature differential between indoor and outdoor is 0ºC; in other words, ventilation only occurs when the outdoor is cooler than the indoor. If the indoor temperature gets down to 19ºC, ventilation is shut off too. It was assumed that natural ventilation can only occur when occupants are at home to open the windows. Values corresponding to a clear uncoated glass were set for window's transmittance and reflectance of both solar radiation and visible light.

An *absorptance* value of 30% (light colour) was selected for this study for both walls and roof, in line with the Brazilian building code [64], as it is recognized as one of the most effective and economical strategy for reducing indoor temperatures in hot-humid climates [65].

An exterior *wall thermal transmittance* value of 2 W/m²K was used, based on the Brazilian building code, acceptable for an absorptance value of 0.30—well below the 0.6 threshold—applicable to climatic zones 3-8 (similar to San Pedro Sula). For instance, in tropical countries like Guadeloupe, Martinique, and Guyana (French overseas territories) this value must be less than 2 W/m2 K [66] and in Jamaica's climatic zone 1 less than 3 W/m2 K [67].

Based on the Brazilian building code climatic zone 7 [68] (similar to San Pedro Sula), a *wall thermal mass* value of 130 kJ/m2K, approximately representing a typical wall configuration in Honduras, such as the concrete block or solid brick with exterior and indoor mortar plaster [22]. A *roof thermal transmittance* value of 2 W/m²K was also used, based on the Brazilian building code, acceptable for an absorptance value of 0.30, applicable to climatic zones 7 and 8 (similar to San Pedro Sula). In other tropical countries, this value is stricter. In Philippines it must be lower than 1.5 W/m²K [69]; in Singapore and India [70,71], lower than 1.2 W/m²K; in Jamaica, lower than 1.08 W/m²K [67]. The value used for the simulations approximately represents a typical uninsulated roof configuration in Honduras, such as the sloped aluminum-zinc sheets over a metallic structure, with a gypsum board ceiling and an air gap for electrical ducting [22]. As it is assumed that an aluminium-zinc sheet is used for roof, a low *roof thermal mass* value was assumed of 10 kJ/m²K.

The *air infiltration* value was set to reflect a leaky building, acknowledging that in such contexts, air leakage is typically neither controlled during construction nor specified during the planning phase. *Occupation schedules* and *lighting schedules, power density* and *radiant fraction* were defined based on the Brazilian building code [72] assuming that in the context of Honduras no study exists on this regard, and that the occupation and lighting profiles defined for the Brazilian context might be similar to other LAC countries such as Honduras.

### 2.1.2. Thermal comfort and overheating calculation

In this study, an adaptive upper thermal comfort temperature limit ($t_{upper\_80\%\_acceptability\_limit}$) was adopted for calculating the *indoor overheating hours* (IOH)—a measure of the time during which indoor operative temperature ($t_{op}$) exceed an upper thermal comfort limit in all thermal zones of a specific building. IOH is similar to the *Exceedance Hours* [73] or *Hours of exceedance* [74] metrics; however, it focuses on assessing heat stress in hot conditions. The selection of the adaptive approach is based on field experiments that have shown that in occupant-controlled naturally conditioned spaces the subjective notion of comfort is influenced by the occupants' thermal experiences, preferences, expectations, and availability of control [75,76].

As defined in Equation 1, the upper 80% acceptability limit was selected to allow a lower standard of thermal comfort as a less strict approach according to ASHRAE 55:

$$t_{upper\_80\%\_acceptability\_limit} = 0.31\overline{t_{pm(out)}} + 17.8 + 3.5$$

*Equation 1*

The prevailing mean outdoor temperature ($\overline{t_{pm(out)}}$) was calculated as defined in Equation 2. In this equation the mean daily temperature for the previous day is represented with $t_{e(d-1)}$, and the mean daily temperature for the day before that $t_{e(d-2)}$, and so on. The α was set to 0.9 since ASHRAE 55 suggests it could be more appropriate for climates in which synoptic-scale (day-to-day) temperature dynamics are relatively minor, such as the humid tropics where San Pedro Sula is located [75]:

$$\overline{t_{pm(out)}} = (1 - \alpha)[t_{e(d-1)} + \alpha t_{e(d-2)} + \alpha^2 t_{e(d-3)} + \alpha^3 t_{e(d-4)} + \dots]$$

*Equation 2*

As defined in Equation 3, this study calculates the IOH weighted by the floor area of each individual thermal zone, as follows:

$$IOH = \frac{\sum_z[(\sum_t h_{z,t}) \cdot a_z]}{\sum_z H_z \cdot \sum_z A_z} \times 100\ \%$$

*Equation 3*

where $IOH$ represents the percentage of indoor overheating hours, $h_{z,t}$ represents the hours of exceedance in zone $z$ at timestep $t$ (i.e. 1 hour); where $h_{z,t} = 1$ only if $t_{op}$ in zone $z$ exceeds the adaptive upper thermal comfort threshold ($t_{upper\_80\%\_acceptability\_limit}$) at timestep $t$; $a_z$ is the floor area of zone $z$; $H_z$ is the sum of all occupied hours in all zones $z$; and $A_z$ is the sum of all floor areas of zones $z$. Weighting in the floor area to the IOH calculation gives a more accurate measure of when floor areas of the thermal zones vary significantly.

*2.1.3. Daylight performance calculation*

In this study, different dynamic annual daylight performance metrics were evaluated so to study its association with IOH, as they have the potential of giving more representative results for the daylight indoor performance, since the sun and sky conditions are constantly changing with inherent daily and seasonal variations dependent upon local climate conditions [77]. They are described as follows:

- *Useful Daylight Illuminance* (UDI): it represents the percentage of occupied hours during which indoor illuminance remains within recommended thresholds. These thresholds, typically ranging from 100 to 3000 lux, are based on research in domestic settings that identifies optimal daylighting for tasks [78]. UDI not only assesses the adequacy of daylight for activities but also indicates the potential for excessive daylight, which can lead to occupant discomfort and unwanted solar heat gain. Essentially, UDI provides a straightforward method for quantifying both daylight availability and solar exposure using a single evaluative framework [79].
- *Daylight Autonomy* (DA): It represents the percentage of annual daytime hours that a given point in a space exceeds a specified illumination level, with a threshold of 300 lux in this case, indicating how often artificial lighting can be avoided in a workplace [80,81].
- *Continuous Daylight Autonomy* (cDA): Similar to DA, it accounts for values below the illuminance threshold by partially contributing to the final percentage. For example, with a threshold of 300 lux, a sensor reading of 300 lux would contribute 100%, while 150 lux would contribute 50% [81].
- *Spatial Daylight Autonomy* (sDA$_{50\%}$): it represents a measure of daylight illuminance sufficiency for a given area [77]. The goal of this metric is to achieve a minimum of 300 lux on a surface located 0.75 meters above the floor area for at least 50% of the annual occupied hours.

The daylight analysis was conducted over a period from 6 a.m. to 7 p.m. (13 hours), accounting for variations in sunrise and sunset throughout the year in San Pedro Sula. To enhance the precision of the daylight simulation results, a sensor grid size of 0.20 meters was specified in Honeybee/Radiance, where a grid-based recipe generated the grid from the floor plans of the Honeybee rooms [55].

*2.2. Overall Desirability calculation*

Many optimization problems involve the analysis of several responses, such as in this study. A useful approach to optimization of multiple responses is the *desirability function* technique. The desirability function translates the functions $y_i$ to a common scale, and then, it blends these functions together using the geometric mean so to obtain an overall desirability response [82]. In this research, the Overall Desirability (D) expresses both IOH (say $y_1$) and UDI (say $y_2$) responses.

The general approach of this technique is to convert each response $y_i$ into an individual desirability function $d_i$ that varies over the range $0 \leq d_i \leq 1$, where if the response $y_i$ is at its goal or target, then $d_i = 1$, and if the response $y_i$ is outside an acceptable region, $d_i = 0$. Then, the objective is to maximize the D by calculating the geometric mean, where $D = (d_1 d_2 \cdots d_m)^{1/m}$, where there are $m$ responses. D will be zero if any of the individual responses is undesirable. In this study $m = 2$, as the study only has two $y_i$ responses. Three types of individual desirability functions can be structured, depending on the type of optimization goal: *maximization, minimization* or *target*.

In this research, the individual desirability function for $y_1$ is formulated with the aim of *minimizing* IOH. Specifically, $y_1$ is considered to be desirable when equal to the objective or target value $T$, set at $T = 0$, and undesirable when it exceeds the upper limit $U$, defined as $U = \max\{s_1, s_2, \ldots, s_n\}$, where $s_1, s_2, \ldots, s_n$ represents the results of $n$ simulation runs. A value of 0 means that indoor temperatures never go above that comfort limit. This *minimization* function is expressed in Equation 4 as follows:

$$d_1 = \begin{cases} 1 & y_1 < T \\ \left(\dfrac{U - y_1}{U - T}\right)^r & T \leq y_1 \leq U \\ 0 & y_1 > U \end{cases}$$

<div align="right"><i>Equation 4</i></div>

Additionally, the individual desirability function for $y_2$ is formulated with the aim of *maximizing* UDI. Specifically, $y_2$ is considered to be desirable when equal to the target value $T$, set at $T = 100$, and undesirable when it fails below the lower limit $L$, defined as $L = \min\{s_1, s_2, ..., s_n\}$, where $s_1, s_2, ..., s_n$ represents the results of $n$ simulation runs. A value of 100% represents the highest achievable value for UDI. This *maximization* function is expressed in Equation 5 as follows:

$$d_2 = \begin{cases} 0 & y_2 < L \\ \left(\dfrac{y_2 - L}{T - L}\right)^r & L \leq y_2 \leq T \\ 1 & y_2 > T \end{cases}$$

<div align="right"><i>Equation 5</i></div>

In both $d_1$ and $d_2$, the weight $r$ was set to a constant value of 1 so that the desirability function is linear. This choice ensures that values less than 1 do not increase the difficulty of meeting the criteria, and values greater than 1 do not make it easier. In RStudio [59], this optimization technique was conducted by using the *desirability* package, where *dMin()*, *dMax()* and *dOverall()* functions were used for setting the targets and upper/lower limits for the minimization (smaller-is-better) function, maximization (larger-is-better) function and combined desirability function, respectively; and the *predict()* function for translating the $d_1$ and $d_2$ responses into D [83].

### 2.3. Screening

As the number of *k* factors in a $2^k$ full factorial design increases, the number of runs required for a complete replicate of the design rapidly may outgrow the resources of any experiment. Hence, fractional factorial designs offer a more resource-efficient alternative to full factorial designs. As explained in 2.3.1, a major use of fractional factorials is in screening experiments—where many factors are considered and the objective is to identify those factors that have large effects. After obtaining the results of the fractional factorial design, the factors identified as important are then investigated more thoroughly in subsequent experiments [84], as explained in 2.3.2.

#### 2.3.1. Fractional factorial design

Based on literature, eight factors were selected: roof overhang depth and WWR, adjusted both by four orientations. These factors are key passive cooling strategies that significantly impact both overheating and daylight in social housing projects. Deeper overhangs and smaller windows reduce solar heat gain during hot periods, which enhances indoor thermal comfort; however, they might reduce indoor daylight. Based on ASHRAE 90.1 standard [85], a WWR of 40% was set, as it is the maximum allowable value recommended for hot-humid climate zones. The lower value of WWR was set to 5%, and not 0%, assuming that all facades need a minimum percentage of window opening for ventilation and daylight purposes. In Honduras, social housing projects are typically built with roof overhangs usually between 50 and 75 cm in depth. Therefore, 50 cm was set as the minimum value in the factorial design. A maximum value of 2.5 meters was chosen, as this depth can provide solar protection and also create a social, semi-outdoor space (e.g., balcony, porch, veranda) [21].

The initial screening steps involved designing a $2^{k-p}$ fractional factorial experiment comprising the 8 two-level factors shown in Table 3, where the {-1, +1} notation is used to code the *low* and *high* levels of each factor, respectively. A $2^8$ full factorial design will result in 256 possible combinations for each IOH and UDI, which is highly resource-intensive, especially considering that every possible combination will simulate all 8760 hours of the year, resulting in more than 4 million hourly results needed to comprehensively calculate IOH and UDI within the full factorial design. For generating the $2^{k-p}$ fractional factorial design *p* needs to be carefully chosen so to ensure that the adequate design combinations are selected for estimating correctly the effects of interest [84]. When creating a fractional factorial design *aliases* may result from the fractioning. For example, in a $2^{3-1}$ design, where *I = ABC*, and *I* is the identity column with which the generator *ABC* is set to be equal, it is impossible to differentiate between *A* and *BC*, *B* and *AC*, and *C* and *AB*, where *A, B,* and *C* are the main effects, and *AB, BC,* and *AC* are two-factor interactions. Considering the latter, in this study a high resolution, *R*, consistent with the degree of fractionation was employed in order to eliminate any *alias effect* in main effects and low-order interactions of the $2^{k-p}$ design. A design is of resolution *R* if no *p*-factor effect is aliased with another effect containing less than *R – p* factors. A Roman numeral subscript is usually used to denote the design resolution.

| Factors | -1 | +1 |
|---|---|---|
| Roof overhang depth, North (meters) | 0.5 | 2.5 |
| Roof overhang depth, South (meters) | 0.5 | 2.5 |
| Roof overhang depth, East (meters) | 0.5 | 2.5 |
| Roof overhang depth, West (meters) | 0.5 | 2.5 |
| WWR, North (%) | 5 | 40 |
| WWR, South (%) | 5 | 40 |
| WWR, East (%) | 5 | 40 |
| WWR, West (%) | 5 | 40 |

*Table 3. Factors and factor levels of the $2_V^{8-2}$ design fractional factorial design*

In this study, a resolution V design was employed so that no main effect or two-factor interaction is aliased with any other main effect or two factor interaction. This resulted in a $2_V^{8-2}$ design fractional factorial design, which corresponds to ¼ fraction of the full factorial design (i.e. 8 factors: A, B, C, D, E, F, G, H), yielding $2^6 = 64$ runs. Considering the following defining relation of *I = ABCDG = ABEFH = CDEFGH*, in this design main effects are confounded with four-factor interactions and two-factor interactions are aliased with three three-factor interactions [84]. In RStudio [59], the fractional factorial design was generated using the *FrF2* package, specifically making use of its *FrF2()* function [86]. After the fractional design was created, the design combinations were saved as a csv file and employed as inputs for thermal comfort and daylight simulations.

### 2.3.2. Stepwise and Lasso regression

Stepwise and Lasso regression were implemented to identify the most significant factors from the pool of 8 factors.

On the one hand, stepwise regression is a method used to select a subset of factors by iteratively adding or removing them based on specific criteria, ensuring that the final model is parsimonious and explanatory [87–89]. In RStudio [59], stepwise regression was implemented using the *stepAIC()* function from the MASS package [90]. Both forward and backward directions were utilized, meaning the algorithm begins with an initial model and explores both adding and removing factors at each step, ensuring that the final model includes only the most significant factors while avoiding overfitting, as suggested by RSM literature [91,92]. The model was evaluated using the Bayesian Information Criterion (BIC), by specifying that $k = \log(n)$, where $k$ is the multiple of the number of degrees of freedom used for the penalty, and the $n$ is the number of samples in the dataset.

On the other hand, Lasso is a regression method that introduces a penalty term that forces the coefficients of the least contributive variables to be exactly zero, keeping only the most significant variables in the final model, and thus, perform variable selection [93]. In Rstudio, the *glmnet()* function within the *glmnet* package was used for conducting the penalized logistic regression with Lasso penalty [94]. To determine the best lambda value, a cross-validation with 3 folds was conducted (*nfolds = 3*). The one standard error rule was applied during cross-validation for model selection (*lambda.1se*).

Considering the deterministic nature of EnergyPlus and Radiance simulation engines, in the subsequent optimization stage using response surface designs, values of statistically non-significant factors according to the stepwise and Lasso regressions are randomized using a normal distribution. This randomization was done using the *rnorm()* function from RStudio's stats package [59].

### 2.4. Optimization

This study followed an RSM approach for optimizing thermal comfort and daylight simultaneously. This study seeks to find the levels of $x_i$ that maximize the overall desirability $y$. The $y$ is a function of the levels of $x_i$, say $y = f(x_1, x_2) + \epsilon$ where $\epsilon$ represents the noise or error observed in the response $y$. The expected response is denoted by $E(Y) = f(x_1, x_2) = \eta$, therefore, the surface represented by $\eta = f(x_1, x_2)$ is called a *response surface*. Considering that in most RSM problems the relationship between the response and the independent variables in unknown, the first step in RSM is to find a suitable approximation for the true functional relationship between $y$ and the set of independent variables. If the response is well modelled by a linear function of the independent variables, then the approximating function is the *first-order model*

$$y = \beta_0 + \beta_1 x_1 + \beta_2 x_2 + \cdots + \beta_k x_k + \epsilon$$

*Equation 6*

If there is curvature in the system, then a polynomial of higher degree is used, such as the *second-order model*

$$y = \beta_0 + \sum_{i=1}^{k} \beta_i x_i + \sum_{i=1}^{k} \beta_{ii} x_i^2 + \sum\sum_{i<j} \beta_{ij} x_i x_j + \epsilon$$

*Equation 7*

The model parameters can be estimated most effectively if proper experimental designs are used to collect the data. Designs for fitting response surfaces are called *response surface designs.* In this study this is discussed in subsections 2.4.1 and 2.4.2.

### 2.4.1. Steepest ascent method

RSM is a *sequential procedure*. Often, the current operating conditions is remote from the region of optimum. It can be identified that is far from the optimum because there is little curvature in the system, therefore, a first-order model will be appropriate. The objective of the RSM is to lead the experimenter rapidly and efficiently along a path of improvement toward the general vicinity of the *optimum*. Once the region of the optimum has been found a second-order model may be employed. When assuming that a first-order model is an adequate approximation to the true surface in a small region of the x's and when the experimenter is far from the optimum region, the method of steepest ascent is used for moving sequentially in the direction of the maximum increase in the response. Experiments are conducted along the *path of steepest ascent* until no further increase in response is observed. This path is taken through the *centre point* of the region of interest, and the actual *step size* is determined by the experimenter based on process knowledge or practical considerations. Then a new first-order model may be fit, a new path of steepest ascent determined, and the procedure continued. Eventually, the experimenter will arrive to the vicinity of the optimum which is usually indicated by a second-order model with better fit than a first order model [84].

For fitting a first-order model, an *orthogonal first-order design* was used as it allows to ensure that observed changes in the response can be attributed uniquely to one factor without interference from the effects of other factors, and to minimize the variance of the estimated regression coefficients. This design is a $2^k$ factorial design that codes the *low* and *high* levels (i.e., $n_f$ factorial points) of the *k* factors to the usual {+1, -1} levels, respectively. Since the $2^k$ design does not afford an estimate of the experimental error unless some runs are replicated, the design is augmented with several $n_c$ centre points. To generate an orthogonal first-order design for fitting a first-order model, the *cube()* function within the *rsm* package in RStudio was used [95]. This design will have 13 runs, 8 corresponding to the $n_f$ factorial points, and 5 to the $n_c$ centre points. After generating the orthogonal first-order design, generated design combinations are saved as a csv file and employed as inputs for thermal comfort and daylight simulations. Later, outputs from Honeybee are appended back to the dataset in RStudio for implementing the RSM using the *rsm* package in RStudio, specifically making use of the *rsm()* function and its corresponding *FO()* argument in the model formula to specify a first-order response surface.

Contour plots are plotted using the *contour()* function within the *rsm* package in RStudio to visually observe the path of steepest ascent. The direction of the steepest ascent is analysed using the *steepest()* function within the same library.

*2.4.2.Location of the stationary point and characterization of the response surface*

When the experimenter is relatively close to the optimum, a model that incorporates curvature is usually required to approximate the response, in most cases, a second-order model is adequate. When fitting this kind of model, the objective is to find the levels of $x_1, x_2, \ldots, x_k$ that optimize the predicted response. This point, if it exists, will be the set of $x_1, x_2, \ldots, x_k$ for which the partial derivatives $\partial \hat{y}/\partial x_1 = \partial \hat{y}/\partial x_2 = \cdots = \partial \hat{y}/\partial x_k = 0$. This point, say $x_{1,s}, x_{2,s}, \ldots, x_{k,s}$, is called the *stationary point*. The stationary point could represent a *maximum point, minimum point,* or *saddle point*. The most straightforward approach to characterise is to examine the response surface using a *contour plot*. However, a more formal analysis, called the *canonical analysis* [84].

Canonical analysis is used for characterising the response surface, where the model is transformed into a new coordinate system with the origin at the stationary point $\mathbf{x_s}$ and then axes of the system are rotated until they are parallel to the principal axes of the fitted response surface [84]. This transformation results in the fitted model

$$y = \hat{y}_s + \lambda_1 \omega_1^2 + \lambda_2 \omega_2^2 + \cdots + \lambda_k \omega_k^2$$

*Equation 8*

Where $\{\omega_i\}$ are the transformed independent variables, or *canonical variables,* and the $\{\lambda_i\}$ are constants or eigenvalues. Like the path of steepest ascent, for second-order models a canonical path determines a linear path along one of the canonical variables that originates at the stationary point. The nature of the response surface can be determined from the stationary point and the *signs* and *magnitudes* of $\{\lambda_i\}$. If the $\{\lambda_i\}$ are all positive, $\mathbf{x_s}$ is a point of minimum response; if the $\{\lambda_i\}$ are all negative, $\mathbf{x_s}$ is a point of maximum response; and if the $\{\lambda_i\}$ have different signs, $\mathbf{x_s}$ is a saddle point. Furthermore, the surface is steepest in the $\omega_1$ direction for which $|\lambda_i|$ is the greatest.

Variations of the pure maximum, minimum, or saddle point response surfaces exists, such as the Ridge systems. On this regard, in the canonical form of the second-order model (Equation 8), if the stationary point $\mathbf{x_s}$ is within the region of experimentation, and one or more of the $\lambda_i$ be very small (e.g., $\lambda_i \simeq 0$), the response variable becomes insensitive to the variables $\omega_i$ multiplied by the small $\lambda_i$. This condition results in a pronounced elongation in the $\omega_i$ direction, allowing the optimum to be taken anywhere along that line. Such a response surface is referred to as a *stationary ridge*. If the stationary point is far outside the region of exploration for fitting the second-order model and one (or more) $\lambda_i$ is near zero, then the surface may be a *rising ridge*. In this type of ridge system, inferences cannot be drawn about the true surface or the stationary point because $\mathbf{x_s}$ is outside the region where the model has been fitted. However, further exploration is warranted in the $\omega_1$ direction.

For fitting a second-order model, a *central composite design (CCD)* was used as it allows the estimation of a quadratic model by augmenting the prior two-level *2^k* factorial design initially employed for fitting the first-order model [84]. Generally, the CCD consists of a *2^k* factorial design with $n_f$ factorial points that is augmented by adding *2k* axial/star points and $n_c$ centre points. The practical deployment of a CCD arises through sequential experimentation, that is, a *2^k* has been used to fit a first-order model, and when that model exhibits lack of fit, axial/star points are then added to allow the quadratic terms to be incorporated into the model. It is important for the second-order model to provide a reasonable and stable variance of the predicted response at all points that are equidistant from the design centre. The latter is achieved by making the design to be *rotatable*, which is a desirable property when the location of the optimum is unknown prior to running the experiment. A CCD is made rotatable by the choice of $\alpha$, which depends on the number of factorial points in the factorial portion of the design, this is, $\alpha = \sqrt[4]{n_f}$. Additionally, in order for the design to provide also reasonably stable variance of the predicted response it is recommended to have between three and five centre points ($3 \leq n_c \leq 5$). In this study it was fixed to $n_c = 3$ [84].

To generate a CCD for fitting a second-order model in RStudio [59], the *ccd()* function within the *rsm* package was used [95]. In this CCD the $\alpha$ parameter was set to "rotatable". This design will have 17 runs, 8 corresponding to the $n_f$ factorial points, 3 to the $n_c$ centre points, and 6 to the axial/star points. After generating the CCD, generated design combinations are saved as a csv file and employed as inputs for thermal comfort and daylight simulations. Later, outputs from Honeybee are appended back to the dataset in RStudio for implementing the RSM using the *rsm* package in RStudio, specifically making use of the *rsm()* function and its corresponding *SO()* argument in the model formula to specify a second-order response surface. A comparison between the quadratic model and the linear model is made by comparing them in terms of the *Bayesian Information Criterion* (BIC) and the *Akaike Information Criterion* (AIC). If the BIC and AIC metrics are lower in the quadratic model is higher it can be said that it makes a better trade-off between the model fit and its complexity than the linear model. Such metrics are extracted from the RSM models.

For characterising the response surface—locating the stationary point to characterise it as a maximum, minimum or saddle point—both contour plots and canonical analysis are implemented using the *contour()* and *canonical.path()* functions, respectively, within the *rsm* package in RStudio.

### 2.5. Robustness

To evaluate the robustness of the optimal (stationary maximum) point identified by the response surface design, a bootstrap procedure was conducted. This approach allows for the estimation of the variability and confidence intervals of the input factors [96]. First, a seed value of 123 was set to ensure reproducibility of the results. Then, the fitted values were predicted and the residuals were calculated from the fitted model for identifying the optimal point. Subsequently, 1000 bootstrap replications were generated using the *replicate()* function in R. In each replication, a new set of pseudo-observed values was created by adding sampled residuals (with replacement) to the fitted values. The model was then updated with these new pseudo-observed data. The bootstrap samples, initially in coded values, are converted back to the original units using the *code2val()* function. The entire process was conducted in RStudio using the *predict()* and *resid()* functions from the Rstudio's *base* library, and the *sample()* function for resampling the residuals. The 95% confidence intervals were calculated for each parameter using the quantiles of the bootstrap distributions, so to provide a measure of uncertainty around the estimated stationary points.

### 3. Results

The following subsections present the results for each research stage of the RSM described in the Methodology section. DoE 0 corresponds to the screening stage, DoE 1 through DoE 5 pertain to the optimization stage using first-order designs, and DoE 6 represents the optimization stage using second-order designs. This study involved **138 simulation** runs to determine the optimal value of *D* that simultaneously optimizes IOH and UDI. From all daylight metrics, only UDI was found to be statistically correlated with IOH (see Table 4); therefore, focus was given to UDI in the multi-objective optimization process along IOH. Using the laptop specifications detailed in Table 5, all simulations took approximately 30 hours and 8 minutes in total to complete. A summary of the results can be found in that table and Figure 4.

| Variables | Spearman's rho |
|---|---|
| IOH & UDI | -0.380815 ($p$ = .002) |
| IOH & DA | 0.05233516 ($p$ = .681) |
| IOH & cDA | 0.04024725 ($p$ = .752) |
| IOH & sDA$_{50\%}$ | 0.01486546 ($p$ = .907) |

*Table 4. Spearman's rank correlation rho between IOH and daylight performance metrics, including p-values.*

| Design | Center | Type | Runs | Time | Key results |
|---|---|---|---|---|---|
| DoE 0 | | $2_V^{8-2}$ fractional factorial design | 64 | IOH: [23 min]<br>UDI: [17 h, 31 min] | • Roof overhang depth on the south and west (i.e. $x_1$), WWR on the west (i.e. $x_2$), and WWR on the south (i.e. $x_3$) are the most significant factors. |
| DoE 1 | $x_1$: 2.5m,<br>$x_2$: 15%,<br>$x_3$: 40% | $2^2$ orthogonal design,<br>$n_c = 3$ | 11 | IOH: [3 min]<br>UDI: [2 h, 11 min] | • Path of improvement (steepest ascent direction) detected: $x_1$ should increase, $x_2$ and $x_3$ should decrease. |
| DoE 2 | | Steepest ascent path | 12 | IOH: [3 min]<br>UDI: [2 h, 37 min] | Re-centre new experiment at distance (in coded units) of +1.5 [$x_1$: 2.958m, $x_2$: 5.05%, $x_3$: 33.53%]. |
| DoE 3 | $x_1$: 2.958m,<br>$x_2$: 5.05%,<br>$x_3$: 33.53% | $2^2$ orthogonal design,<br>$n_c = 3$ | 11 | IOH: [3 min]<br>UDI: [2 h, 21 min] | • Path of improvement (steepest ascent direction) detected: $x_1$ should increase, $x_2$ and $x_3$ should decrease. |
| DoE 4 | | Steepest ascent path | 12 | IOH: [3 min]<br>UDI: [2 h, 34 min] | Re-centre new experiment at distance (in coded units) of +1.5 [$x_1$: 3.513m, $x_2$: 4.723%, $x_3$: 23.56%]. |
| DoE 5 | $x_1$: 3.513m,<br>$x_2$: 4.723%,<br>$x_3$: 23.56% | $2^2$ orthogonal design,<br>$n_c = 3$ | 11 | IOH: [3 min]<br>UDI: [2 h, 16 min] | • R-squared dropped to 0.2342 and model p-value > .05. Second order design is necessary as the latter may suggest a curvature in the system. |
| DoE 6 | $x_1$: 3.513m,<br>$x_2$: 4.723%,<br>$x_3$: 23.56% | CCD: $2^3$,<br>$\alpha$ points = 6,<br>$n_c = 3$ | 17 | IOH: [5 min]<br>UDI: [3 h, 41 min] | • BIC and AIC are lower than those observed in DoE 5. Second-order design better than first-order one.<br>• All eigenvalues are negative, thus, stationary point identified. |

* For a 11th Gen Intel® Core™ i5-1135G7 @ 2.40GHz 1.38 GHz, RAM 7,74GB, Win10 x64 Processor

*Table 5. Summary of the sequential multi-objective optimization procedure followed in this case study.*

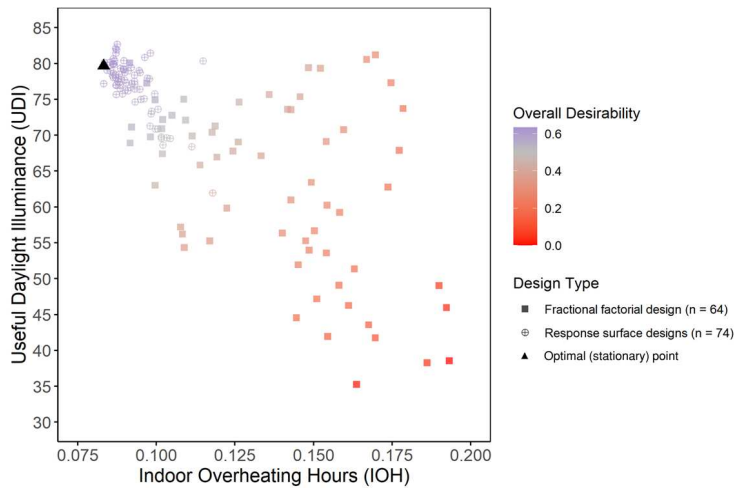

*Figure 4. Scatterplot showing all simulation outputs, colouring each output with its corresponding D.*

### 3.1. Screening

A $2_V^{8-2}$ fractional factorial design (i.e. DoE 0) resulted from specifying a resolution V design with $k = 8$, yielding a significant reduction in the number of combinations: from 256 to 64 for each individual response variable (i.e., IOH and UDI). This reduction accounts for 25% of the full factorial design. The $2_V^{8-2}$ design with coded and uncoded units for each of these experimental runs can be found in Supplementary Material (i.e. *frac_design_2k8_R5.csv* file). For the 64 simulations (experiments) of the fractional factorial design, IOH values oscillated between 9.16% and 19.32% of indoor overheating hours; UDI values between, 35.25% and 81.21%. As explained in Methodology section, these aforementioned values were set as the minimum and maximum values, respectively, for calculating the *D* as they represent the natural behaviour of the response variables given the minimum and maximum values set for the input factors. IOH and UDI simulation outputs from the fractional factorial design, transformed into *D* values, ranged from **0 to 0.603.** Statistically significant correlations, as shown in Figure 5, indicate that a decrease in IOH corresponds with an increase in UDI. Additionally, as *D* increases, IOH decreases while UDI rises. As the roof overhang depth increases in the south, IOH decreases while UDI and D increase. In the west, a deeper roof overhang leads to a decrease in IOH and an increase in *D*. In the south and west, increasing the WWR causes UDI and *D* to decrease. In the east, a higher WWR reduces both IOH and UDI. However, while an increase in WWR in the east raises IOH, the same increase in the west leads to a decrease in IOH. The WWR on the north side and the roof overhang depth on the north and east sides show no significant association with any of the response variables.

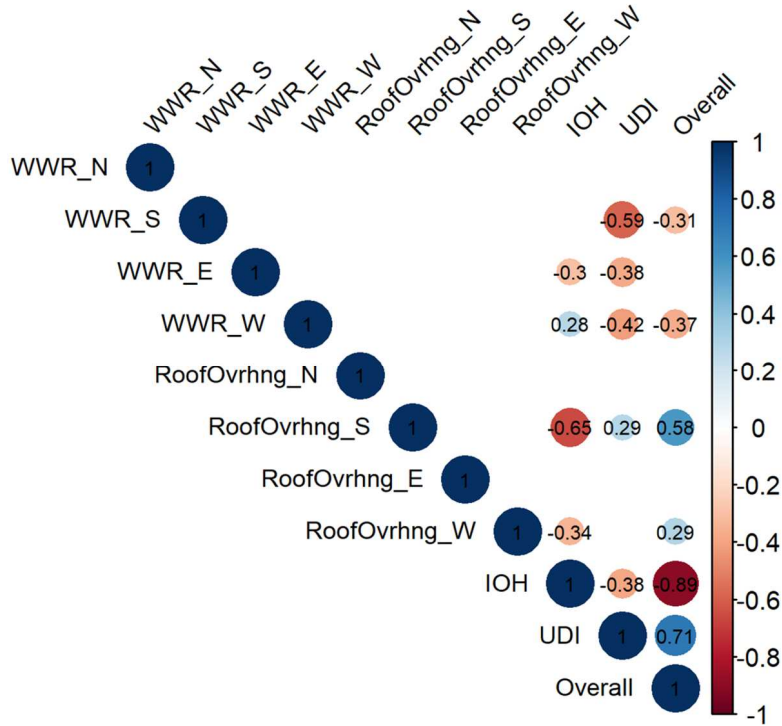

*Figure 5. Correlogram showing the Spearman's correlation rho between factors and response variables (IOH, UDI, D). Note: Only statistically significant correlations (p-value < .05) are displayed.*

Stepwise and Lasso regression identified the most contributive factors affecting $D$. As shown in Table 6, both techniques identified four out of the eight factors as having a statistically significant impact. According to both regression techniques, the factor with highest effect on the $D$ is the roof overhang depth on the south side, and the factor with lowest (but significant) effect is the roof overhang depth on the west side. On the south and west orientations, higher roof overhang depth increases the $D$, but higher WWR decreases the $D$.

| Model parameters | Stepwise regression Estimate (*p*-value) | *Lasso regression* Estimate |
| --- | --- | --- |
| (Intercept) | .33336 ($p < .001$) | .263 |
| Roof overhang depth [South] | .08659 ($p < .001$) | .130 |
| WWR [West] | -.05728 ($p < .001$) | -.072 |
| WWR [South] | -.04863 ($p < .001$) | -.054 |
| Roof overhang depth [West] | .04286 ($p < .001$) | .043 |

*Table 6. Regression results using simulation outputs of the $2_V^{8-2}$ fractional factorial design.*

As the RSM follows a sequential procedure, these four factors were chosen for the optimization stage. However, for simplification purposes, they were condensed down to three factors in the following way: south & west roof overhang depth ($x_1$, positively associated with the $D$), west WWR ($x_2$, negatively associated with the $D$), and south WWR ($x_3$, negatively associated with the $D$). As previously explained in 2.3.2 subsection, factors with negligible effect on the $D$ (i.e., WWR and roof overhang depth on both the north and east) were assigned random values following a normal distribution with a mean value (i.e. WWR: 15%; roof overhang: 0.50 meters) and a standard deviation of 2.

### 3.2. Optimization

Opposite to the screening stage where only two levels $\{+1, -1\}$ are evaluated per factor, in the RSM optimization stage first-order designs evaluate three levels $\{+1, 0, -1\}$ and second-order designs evaluate five levels $\{+\alpha, +1, 0, -1, -\alpha\}$. A total of 74 simulations were conducted—57 using first-order designs and 17 using second-order designs—to identify the optimal value for each input factor that optimizes the $D$. Sections 3.2.1 and 3.2.2 detail the sequential, step-by-step process of the response surface designs. This process identified the optimal point at DoE 6.

*3.2.1. Fitting first-order models*

**Design of Experiment 1 (DoE 1): initial experiment.** The initial operating conditions of the experiment were established without any prior knowledge of the location or distance to the optimum. For the DoE 1, the starting operating conditions (i.e. central points) of the first-order design were set at a roof overhang depth of 2.5 meters on the south and west sides, 15% of WWR on the west, and a 40% of WWR on the south. The results of the DoE 1 yielded *D* centre point values ranging from **0.4929 to 0.5030**, which are far from the optimum *D* value of 1. In their original metrics, these correspond to IOH values between 10.15% and 10.21% and UDI values between 68.63% and 69.78%. The first-order model indicated that decreasing WWR on the west and south sides while increasing the roof overhang depth would likely improve the response. The high R-squared value and significant model p-value suggested that no curvature was present, warranting an exploration along the steepest ascent path.

**Design of Experiment 2 (DoE 2): exploring the path of steepest ascent.** The aim of implementing the steepest ascent method is to find the highest point along this path, and centre the next experimental design there. Since the response surface model suggested a path of improvement for the x's, the steepest ascent method was implemented. Plotted results suggest that the centre for the next experimental design should be at a distance of about 1.5 in coded units, where the *D* reached its highest point with a value of **0.6030** (see Figure 6)**.** In the original metrics, this highest point corresponds to an IOH value, in percentage, of 8.98%, and an UDI value, in percentage, of 79.24%. At this distance, the simulation input value of the roof overhang depth on the south and west sides is 2.958 meters, the simulation input value of the WWR on the west side is 5.05%, and the simulation input value of the WWR on the south side is 33.53%.

**Design of Experiment 3 (DoE 3): relocating the experiment.** DoE 3 is basically the same as the DoE 1, however, with the new centre points suggested by the DoE 2. Results of the DoE 3 yielded *D* centre point values between **0.5879 and 0.6113**, which are closer to the optimum (i.e. *D* value of 1) compared to the DoE 1. In their original metrics, these are IOH values that oscillated, in percentage, between 8.96% and 9.05%, and UDI values, in percentage, between 77.33% and 80.42%. The steepest ascent direction yielded by the response surface model built from the DoE 3 suggested decreasing the WWR value on the west and south sides (i.e. $x_2$ and $x_3$, respectively) while increasing the roof overhang depth on the south and west sides (i.e. $x_1$). The R-squared value decreased compared to DoE 1, and the model p-value worsened to *p* = .0285. However, this still does not indicate curvature in the system. As a result, the direction of steepest ascent was reassessed.

**Design of Experiment 4 (DoE 4): exploring again the path of steepest ascent.** Since the first-order model built from the DoE 3 suggested a path of improvement for the x's, the steepest ascent method was implemented again. Plotted results suggest that the centre for the next experimental design should be at a distance of about 1.5 in coded units, where the *D* reached its highest point with a value of **0.6165** (see Figure 6)**.** In the original metrics, this highest point corresponds to an IOH value, in percentage, of 8.66%, and an UDI value, in percentage, of 79.87%. At this distance, the simulation input value of the roof overhang depth on the south and west sides is 3.513 meters, the simulation input value of the WWR on the west side is 4.723%, and the simulation input value of the WWR on the south side is 23.56%.

**Design of Experiment 5 (DoE 5): relocating experiment again.** DoE 5 is basically the same as the DoE 1 and DoE 3, however, with the new centre points suggested by the DoE 4. Results of the DoE 5 yielded *D* centre point values between **0.6049 and 0.6287**, closer to the optimum (i.e. *D* value of 1) when compared to previous DoEs. In their original metrics, these are IOH values that oscillated, in percentage, between 8.63% and 8.74%, and UDI values, in percentage, between 78.07% and 81.97%. The R-squared value decreased further compared to DoE 3, and the model p-value increased to a *p* > .05. This undermines the reliability of the steepest ascent direction suggested by the first-order model from DoE 5, indicating potential curvature in the system. Consequently, a second-order experiment with the same central point was necessary. This experiment should be compared to DoE 5 using AIC and BIC metrics.

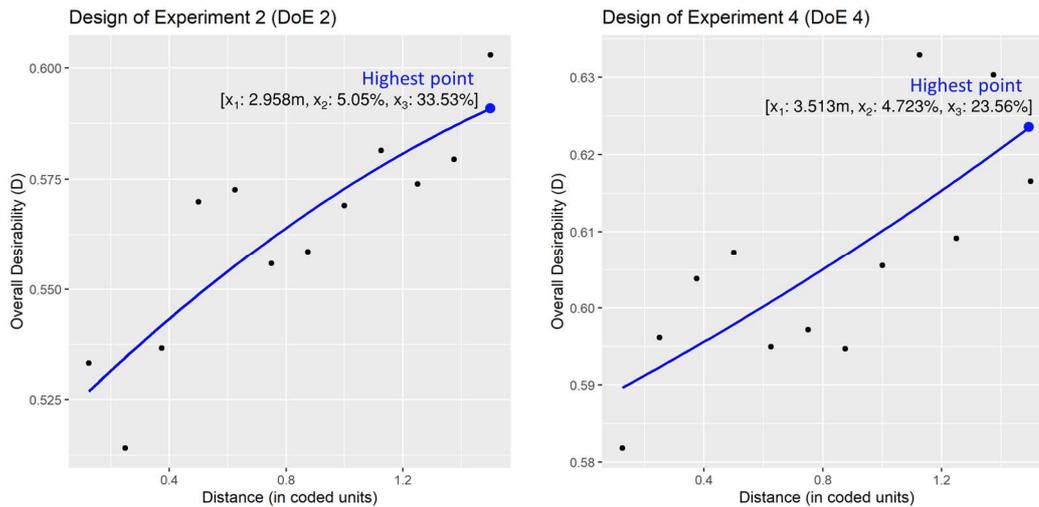

*Figure 6. The path of steepest ascent for DoE 2 and DoE 4. Each path highlights the distance, measured in coded units, at which the highest D value was achieved.*

### 3.2.2. Fitting second-order models

**Design of Experiment 6 (DoE 6): fitting second-order experiment.** DoE 6 is basically the same as the DoE 5, however, instead of having three levels per factor it has five, two levels more corresponding to the star/axial points of the CCD. Results of the DoE 6 yielded *D* centre point values between **0.6201 and 0.6257.** In their original metrics, these are IOH values that oscillated, in percentage, between 8.63% and 8.71%, and UDI values, in percentage, between 80.42% and 81.44%. As mentioned before, results of DoE 5 need to be compared with results of DoE 6 in order to determine which model gives a better prediction. BIC and AIC values of the second order design, -99.28 and -90.82, respectively, exhibits a better fit than the BIC and AIC values of the first order design, -59.09 and -57.11, respectively. Therefore, DoE 6 has a better fit. The stationary point of the second-order model—roof overhang on the south and west sides of 3.78 meters, WWR on the west of 3.76%, and WWR on the south of 29.34%—can indeed be characterised as a maximum point since all eigenvalues are negative ($x_1$: -0.002745706, $x_2$: -0.003623011, $x_3$: -0.021416661), as illustrated in Figure 7. The latter, thereby, means that the optimal *D* was found. Consequently, the optimal IOH and UDI. The quadratic model explains approximately the 92% (R-squared = 0.9241, $p$ = .004) of the variability of the *D*, and has a non-significant lack of fit ($p$ = .050). It has significant first-order effects ($p$ = .004) and quadratic effects ($p$ = .002), and non-significant interaction effects ($p$ = .360).

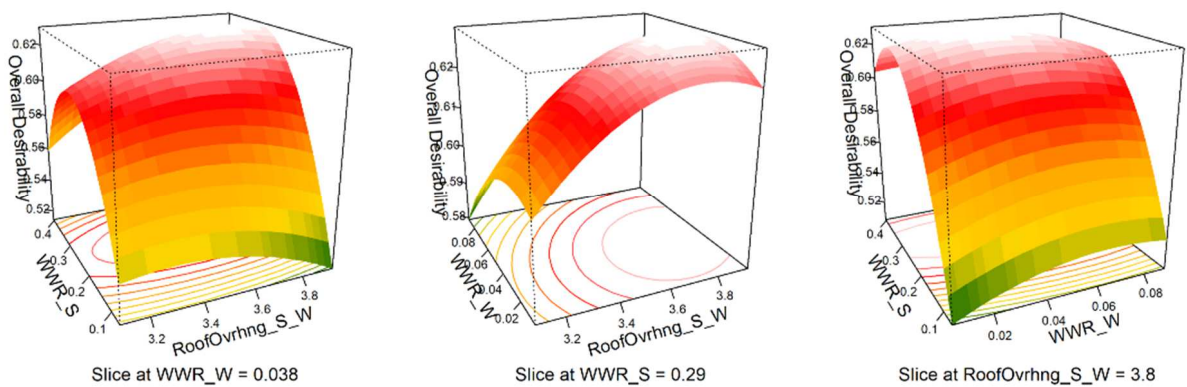

*Figure 7. 3D surface plots illustrating RSM results for DoE 6.*

This stationary, maximum point was evaluated, by conducting one last additional simulation (i.e., experiment 139), thus, validating that it is indeed optimal (see Figure 8). The simulation run output of IOH and UDI yielded a *D* value of **0.6247 (IOH: 8.33%; UDI: 79.67%)**.

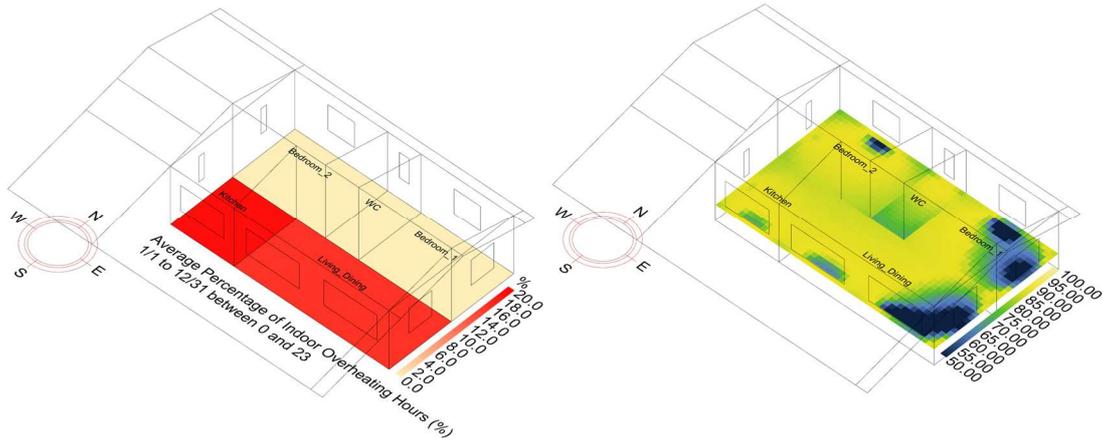

*Figure 8. 3D Rhino/Honeybee visualization of the housing typology, illustrating the optimal IOH (left) and UDI (right): experiment (simulation) 139.*

### 3.3. Robustness

Figure 9 illustrates the confidence regions for the stationary point identified as a maximum point in the second-order model derived from the DoE 6 were identified. According to the bootstrap procedure, the optimal roof overhang depth for the south and west sides has a 95% confidence interval of [3.47, 3.94] meters. For the WWR at the west side, the optimal point has a 95% confidence interval of [1.54, 5.59] %. The optimal WWR at the south side has a 95% confidence interval of [26.19, 31.78] %.

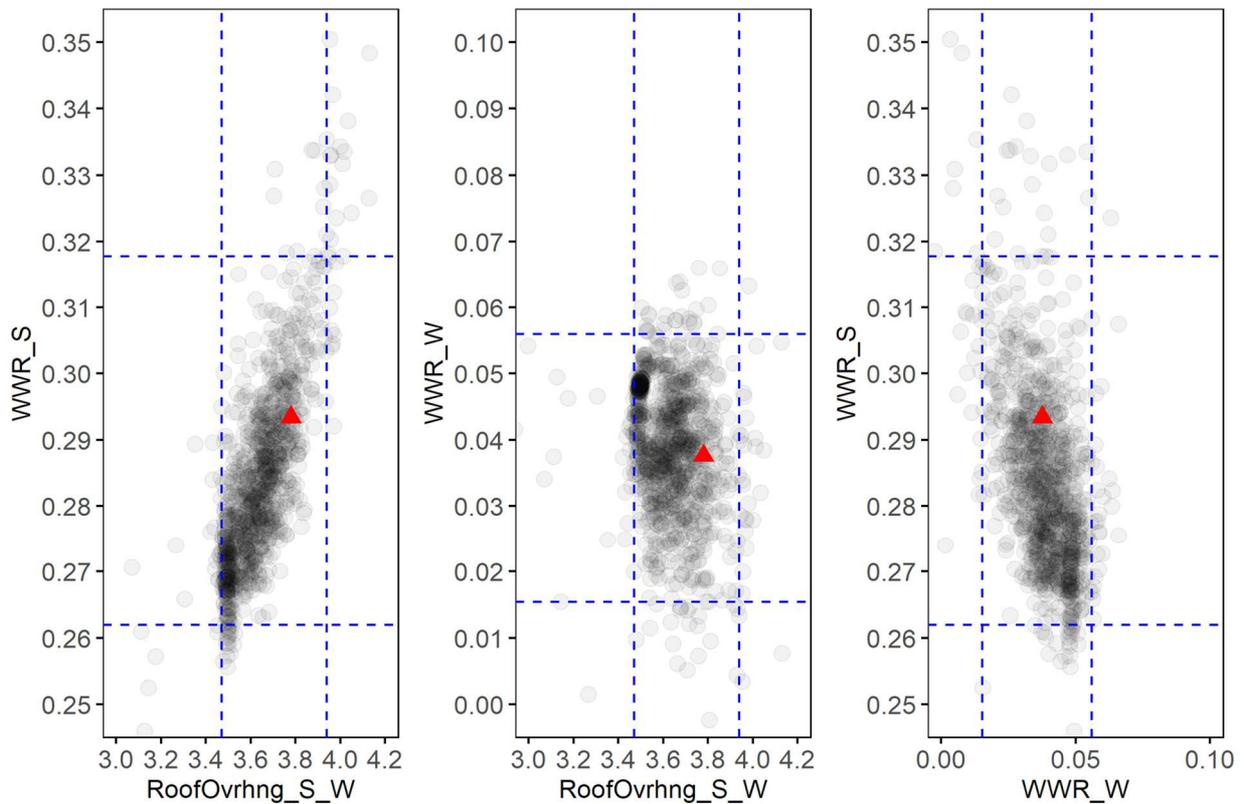

*Figure 9. Bootstrap results showing the 95% confidence regions (constrained by blue dashed lines) of the optimal point (red triangle) identified through RSM.*

## 4. Discussion

*The identified optimal values are validated by the presence of all negative eigenvalues and further assessed by bootstrap analysis.* Results are consistent with existing literature [97–99], allowing lower values of WWR on the west side in comparison to the south side; and long roof overhangs on both facades. This study offers unique insights by providing both the optimal values and their narrow confidence regions: approximately ±2% for the WWR on the west façade, ±2.75% for the WWR on the south façade, and ±0.25 meters for the roof overhang depth.

*This study is the first to apply an RSM-based multi-objective optimization procedure specifically for mitigating overheating and enhancing daylight performance.* A key advantage of using RSM for multi-objective optimization is the efficiency it offers, requiring fewer simulation runs as it is based on DoE and providing clear control over the optimization process, especially given the time-intensive nature of thermal comfort and daylight annual simulations. Only 138 simulations for each individual response (~30 hours for 276 runs) were needed using a basic office laptop, demonstrating RSM's ability to maintain control over the optimization process while continuously generating new data as it converges on the optimal point. Unlike genetic algorithms, RSM offers a transparent and computationally-effective optimization process with fewer simulations. On this regard, a study referenced in 1.3 conducted 12000 simulations to train a CatBoost regressor model for predicting daylight metrics (e.g. UDI, sDA), and then used the surrogate model as the objective for the NSGA-III optimization algorithm [51]. Another study used simulation sample sizes of 500, 1000, 1500, and 2000, and combined BO-XGBoost with NSGA-II for the multi-objective optimization of energy intensity, thermal comfort and daylight [54]. While none of them specify the total duration of the process, they emphasize the substantial computational resources required.

*The optimal point identified by RSM does not rule out the potential for further refinement by considering additional measures.* For instance, Figure 8 shows that IOH levels are higher in south-facing rooms compared to north-facing rooms. To further reduce IOH in these areas, strategies such as increasing the roof U-value or adding a double roof—both unrelated to daylight—could be implemented. Also, in terms of UDI, it shows a blue region on the east façade, likely associated with high glare levels (i.e. low UDI values). The performance could be further improved by adding in the east, for instance, solar protection. However, it is worth noting that the stepwise regression model suggests that changes to the WWR or roof overhang depth on the east side do not significantly improve the *D* in this tropical dwelling typology, as it is a factor significantly uncorrelated with *D* (see Figure 5). This underscores the importance of screening before applying RSM, as emphasized in the literature, to reduce overfitting and simplify the model. Following recommendations from the RSM literature [91,92], in this study, p-values from stepwise regression, along with Lasso regression—which forces the coefficients of the least contributive variables to zero through a penalty term— were used during the screening stage to identify which factors advanced to the RSM optimization phase.

Additionally*, the optimization process was context-specific, focusing on the initially specified passive thermal and daylight performance factors (e.g., WWR and roof overhang on all orientations)* indicated in Table 3. On the one hand, other passive strategies that could impact indoor overheating, such as low roof U-values or low-albedo roof values, were fixed without any variation. These measures could significantly reduce indoor cooling or thermal needs [100]. On the other hand, the optimization did not account for variations in visible light transmittance (VLT) or visible light reflectance (VLR) of window glass. VLT impacts natural light and solar radiation entry, affecting artificial lighting needs, while VLR influences glare and the amount of light reflected back into the environment [101,102]. Unfortunately, window manufacturers or sellers in the Honduran low-income context often do not provide these specifications, and they typically fall outside the expertise of most architects. The study focused on widely recognized passive measures, such as WWR and roof overhangs, which architects commonly identify as key factors in managing overheating and daylight.

*Future studies should further explore the relationship between overheating and daylight, as this is the first study of its kind to examine this association.* IOH correlates significantly with UDI (see Table 4) very likely since both deal with excessing solar radiation and daylight, which can lead to occupant discomfort and unwanted solar heat gain. By focusing this case study on IOH and UDI—given that no significant association was found between IOH with DA, cDA and sDA—the optimization process was streamlined for better interpretation of results. Including other daylight metrics that do not correlate well with IOH could introduce unnecessary complexity and dilute the optimization focus without adding significant value.

*Incorporating calibrated digital models would further enhance the research by enabling a more detailed evaluation of the methodology's robustness*, especially since this study relied on a parametric approach. To enhance the generalizability of these findings, future research should broaden the scope by applying the methodology to a diverse set of randomly selected social housing digital typologies, rather than concentrating on a single model. Nonetheless, literature suggests that single-family dwellings experience greater overheating compared to apartment-type dwellings [103]. A previous simulation-based study conducted in San Pedro Sula, involving nearly 4,000 thermal simulations [21], found that for an apartment-type dwelling, the best IOH was 20.4%. In contrast, the current study reveals under the same climate scenario that the optimal IOH point for this single-family typology is 8.23% (i.e., percentage of annual hours requiring air-conditioning).

## 5. Conclusions

This case study demonstrates how applying Response Surface Methodology (RSM) with desirability functions can optimize passive measures—such as window-to-wall ratio (WWR) and roof overhang depth—early in the building design process with minimal simulation (experimentation) effort. By leveraging advanced statistical methods and building simulation models this research highlights how a computationally-efficient multi-objective approach can effectively reduce the Indoor Overheating Hours (IOH) and enhances Useful Daylight Illuminance (UDI), thereby reducing the reliance on mechanical cooling and glare protection systems. The robustness of this method was validated through 1,000 bootstrap replications, providing 95% confidence intervals (regions) for the optimal solutions. This study offers valuable insights for housing policymakers seeking to mitigate climate impacts on indoor environments through cooling/daylight passive strategies.

**Acknowledgements**

The authors thank the Institute of Data Science and Artificial Intelligence (DATAI) of the University of Navarra for providing all the necessary facilities for this work.

**Author contributions**

J.G.S. conducted the analysis, interpreted the results, and wrote the manuscript, while J.L.F. assisted with results interpretation and manuscript revision.

**Competing interests**

The authors declare no competing interests.

**Additional information**

Supplementary material related to this article (e.g. code, Grasshopper simulation files) can be found at the following Github repository: https://github.com/juan-gamero-salinas/rsm-thermal-daylight-optimization